\documentclass[prb,twocolumn,showpacs,preprintnumbers,amsmath,amssymb,floatfix]{revtex4}

\usepackage{graphicx}
\usepackage{bm}
\usepackage{txfonts}
\usepackage{booktabs}
\usepackage{color}
\usepackage{multirow}

\newcommand{\cp}{\ensuremath{c_{p}}}
\newcommand{\cel}{\ensuremath{c_{\rm el}}}
\newcommand{\cph}{\ensuremath{c_{\rm ph}}}
\newcommand{\Dcel}{\ensuremath{\Delta c_{\rm el}}}
\newcommand{\gn}{\ensuremath{\gamma_{\rm n}}}
\newcommand{\gs}{\ensuremath{\gamma_{\rm s}}}
\newcommand{\gres}{\ensuremath{\gamma_{\rm res}}}
\newcommand{\kB}{\ensuremath{k_{\rm B}}}
\newcommand{\NA}{\ensuremath{N_{\rm A}}}
\newcommand{\TD}{\ensuremath{{\it \Theta}_{\rm D}}}
\newcommand{\Hc}{\ensuremath{H_{\rm c}}}
\newcommand{\Hsc}{\ensuremath{H_{\rm sc}}}
\newcommand{\Tc}{\ensuremath{T_{\rm c}}}

\newcommand{\Hcz}{\ensuremath{H_{\rm c2}}}
\newcommand{\Hcd}{\ensuremath{H_{\rm c3}}}
\newcommand{\EF}{\ensuremath{E_{\rm F}}}
\newcommand{\kF}{\ensuremath{k_{\rm F}}}
\newcommand{\vF}{\ensuremath{v_{\rm F}}}
\newcommand{\me}{\ensuremath{m_{\rm el}}}
\newcommand{\Vmol}{\ensuremath{V_{\rm mol}}}

\newcommand{\kGL}{\ensuremath{\kappa_{\rm GL}}}
\newcommand{\HT}{$H$\,--\,$T$}
\newcommand{\mOc}{\ensuremath{\muup\Omega{\rm cm}}}

\begin{document}

\title{Specific heat and electronic states of superconducting boron-doped silicon carbide}

\author{M.~Kriener and Y.~Maeno} 
\affiliation{Department of Physics, Graduate School of Science, Kyoto University, Kyoto 606-8502, Japan}

\author{T.~Oguchi}
\affiliation{Department of Quantum Matter, Hiroshima University, Higashihiroshima 739-8530, Japan}

\author{Z.-A.~Ren, J.~Kato, T.~Muranaka, and J.~Akimitsu}
\affiliation{Department of Physics and Mathematics, Aoyama-Gakuin University, Sagamihara, Kanagawa 229-8558, Japan}

\date{\today}

\begin{abstract}
The discoveries of superconductivity in the heavily-boron doped semiconductors diamond\cite{ekimov04a} (C:B) in 2004 and silicon\cite{bustarret06a} (Si:B) in 2006 have renewed the interest in the physics of the superconducting state of doped semiconductors. Recently, we discovered superconductivity in the closely related ''mixed'' system heavily boron-doped silcon carbide (SiC:B).\cite{ren07a} Interestingly, the latter compound is a type-I superconductor whereas the two aforementioned materials are type-II. In this paper we present an extensive analysis of our recent specific-heat study, as well as the band structure and expected Fermi surfaces. We observe an apparent quadratic temperature dependence of the electronic specific heat in the superconducting state. Possible reasons are a nodal gap structure or a residual density of states due to non-superconducting parts of the sample. The basic superconducting parameters are estimated in a Ginzburg-Landau framework. We compare and discuss our results with those reported for C:B and Si:B. Finally, we comment on possible origins of the difference in the superconductivity of SiC:B compared to the two ''parent'' materials C:B and Si:B.
\end{abstract}



\maketitle

\section{Introduction}
Diamond and silicon are wide-gapped semiconductors / insulators which exhibit indirect energy gaps of about 5.5\,eV (diamond) and 1.1\,eV (silicon). They are well-known for their outstanding physical properties and technical applications, e.\,g.\ the excellent heat conductivity of diamond, its withstanding of high electric fields, or the numerous applications of silicon in semiconductor technologies. It is well-known, too, that the physical properties of these and other semiconductors can be influenced by charge-carrier doping either by donor or acceptor atoms which changes their resistivity many orders of magnitude leading to intriguing properties. Small doping concentrations are widely used in the application of semiconductors. At higher doping levels the systems undergo a semiconductor-to-metal transition above a certain critical doping level, i.\,e.\ charge-carrier concentration, and further doping might even lead to superconductivity. From the theoretical and experimental point of view superconductivity in doped semiconductors is an outstanding issue. The prediction of superconductivity in Ge and GeSi and the suggestion that other semiconductor-based compounds may also exhibit superconductivity at very low temperatures were given by Cohen already in 1964.\cite{cohen64a} Indeed, some examples have been reported so far, e.\,g.\ self-doped Ge$_x$Te,\cite{hein64a} Sn$_x$Te,\cite{hein69a} doped SrTiO$_3$,\cite{schooley65a} or more recently doped silicon clathrates.\cite{kawaji95a,grosche01a,connetable03a} The superconductivity of the doped silicon clathrates is the first example of compounds exhibiting superconductivity in a covalent tetrahedral $sp^3$ network with bond lengths similar to those in diamond.

However, before 2004 (C:B)\cite{ekimov04a} and 2006 (Si:B)\cite{bustarret06a} superconductivity was never reported for diamond and cubic silicon in the diamond structure although there are several studies available concerning hole-doped induced metallicity in carbon and silicon using boron, nitrogen, or phosphorus, e.\,g.\ Refs.~\onlinecite{dai91a}, \onlinecite{bustarret03a} and the references therein. Therefore, it was an important progress to find superconductivity in these compounds upon boron doping, which attracted a lot of interest and stimulated many theoretical and experimental studies in the last four years.\cite{baskaran04a,blase04a,boeri04a,bustarret04a,kwlee04a,takano04a,xiang04a,umezawa05a,yokoya05a,sacepe06a,wu06a,bourgeois07a,shirakawa07a,takano07a} Boron has one partially filled electron less than carbon or silicon and hence acts as an acceptor leading to hole doping. Both compounds are type-II superconductors with \Tc\ values of 11.4\,K (C:B) and 0.35\,K (Si:B). The upper critical fields are $\Hcz\approx 8.7$\,T and 0.4\,T, respectively.\cite{umezawa05a,bustarret06a} 

In order to explain the superconductivity in diamond theoretical studies point towards two different scenarios: (i) it is the result of a simple electron-phonon interaction\cite{blase04a,boeri04a,kwlee04a,xiang04a} and (ii) it is caused by a resonating valence-bond mechanism.\cite{baskaran03a,baskaran06a,baskaran04a} The former model is based on a conventional electron-phonon mechanism where the charge carriers are introduced into intrinsic diamond bands leading to a three-dimensional analog of the two-dimensional superconductor MgB$_2$. The superconductivity is attributed to holes located at the top of the zone-centered $\sigma$-bonding valence bands which couple strongly to optical bond-stretching phonon modes.\cite{boeri04a,kwlee04a} The latter model attributes the superconductivity to holes in the impurity bands rather than in the intrinsic diamond bands.\cite{baskaran04a} With the premise that the doping level in superconducting diamond is close to the Mott limit the randomly distributed boron atoms, i.\,e.\ their random Coulomb potential, may lift the degeneracy of the boron acceptor states leading to a narrow half-filled band from which superconductivity develops. However, spectroscopical studies seem to support the former explanation and rule out the latter suggestion,\cite{kacmarcik05a,sacepe06b} although a complete understanding of the superconducting phase is not yet obtained.\cite{nakamura04b,baskaran06a,wu06a} Recently, a theoretical study suggested the possibility to achieve superconducting transition temperatures on the order of 100\,K in C:B due to the exceptionally high Debye temperature of diamond and under the precondition that the doped boron atoms are ordered.\cite{shirakawa07a} 
\begin{figure}
\centering
\includegraphics[width=8.5cm,clip]{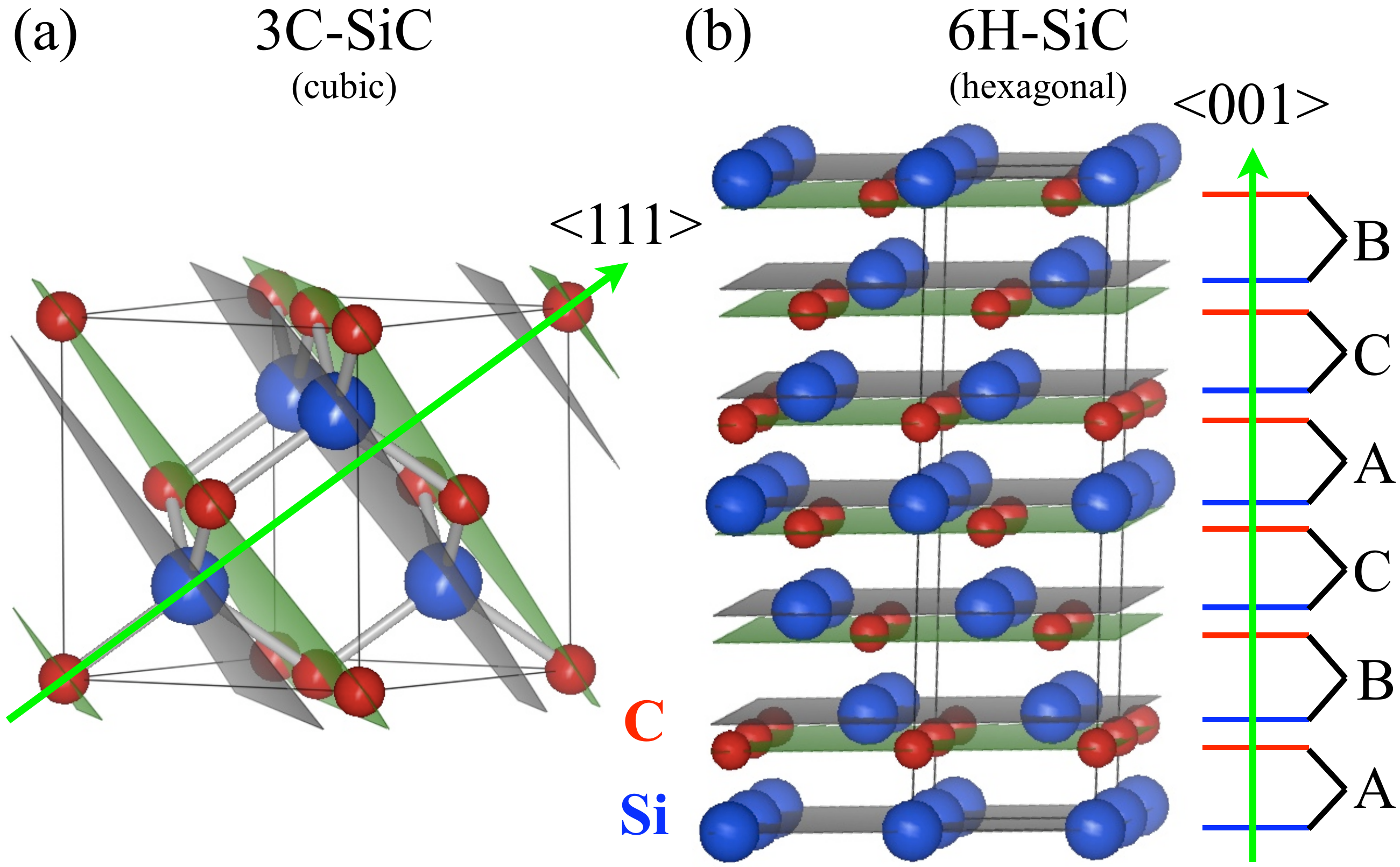}
\caption[]{(color online) (a) Unit cell of diamond-related cubic 3C-SiC. The three bilayers consisting of C and Si layers are emphasized. The stacking sequence is ABC\,--\,\dots The  green arrow denotes the $\left<111\right>$ direction whereas the gray rods refer to the tetrahedral bond alignment of diamond. The cube defines the conventional unit cell of 3C-SiC, which consists of four formula units SiC. (b) Four unit cells of hexagonal 6H-SiC. The cuboid defines one unit cell. The six bilayers of the stacking sequence ABCACB\,--\,\dots are emphasized. The green arrow denotes the $\left<001\right>$ direction. One unit cell of 6H-SiC consists of six formula units SiC. For the drawings the software \textit{Vesta} was used.\cite{Vesta}} \label{structureSiC}
\end{figure}

In Ref.~\onlinecite{ren07a}, we reported the discovery of superconductivity in a closely related system originating from a well-known and widely used semiconductor, namely boron-doped SiC, the stoichiometric ''mixture'' of the two afore discussed ''parent'' materials. SiC is used increasingly for high-temperature, high-power, and high-frequency applications due to its high thermal conductivity, the existence of large band gaps, strong covalent bondings, chemical inertness, or its high tolerance to radiation and heat. Another hallmark of this system is the huge number (about\cite{casady96a} 200) of crystal modifications with cubic (''C''), hexagonal (''H''), or rombohedral (''R'') symmetry of the unit cell.\cite{SiCinfo,casady96a} They are usually referred to as $m$C-SiC, $m$H-SiC, and $m$R-SiC, respectively.
\begin{table}[t]
\centering
\begin{ruledtabular}
\caption{Basic parameters of 3C-SiC and 6H-SiC at room temperature.\cite{landolt} The parameter $V_0$ denotes the volume of the conventional unit cell, $\Vmol=V_{0}\cdot \NA/t$ the molar volume where $t$ is the number of formula units SiC in the unit cell (''f.\,u.\ / unit cell''). In our analysis we will use the average of $\Vmol^{\rm 3C-SiC}$ and $\Vmol^{\rm 6H-SiC}$ because the sample used contains both polytypes; see text.}
\begin{tabular}{lcc}
\toprule
                                   & 3C-SiC                       & 6H-SiC \\ 
                                   & $\beta$-SiC                  & $\alpha$-SiC \\ 
\addlinespace[0.25em]\hline \addlinespace[0.75em]
\multirow{2}{*}{symmetry}          & cubic                        & hexagonal \\ 
                                   & zincblende                   & moissanite-6H\\ 
\addlinespace[0.25em]
space group                        & F$\bar{4}$3m (T$_{\rm d}^2$) & P6$_3$mc (C$^{4}_{6v}$)\\ 
\addlinespace[0.25em]
\multirow{2}{*}{bilayer stacking}  & ABC\,--\,\dots               & ABCACB\,--\dots\\ 
                                   & along $\left<111\right>$     & along $\left<001\right>$\\ 
\addlinespace[0.25em]
 f.\,u.\ / unit cell               & $t=4$                        & $t=6$\\ 
 \addlinespace[0.25em]
\multirow{2}{*}{lattice constants (\AA)} & \multirow{2}{*}{$a_{\rm cub}= 4.3596$} & $a_{\rm hex}= 3.0806$\\ 
                                   &                              & $c_{\rm hex}= 15.1173$\\ 
\addlinespace[0.25em]
$V_0$ (\AA$^3$)                    & 82.859                       & 124.244 \\ 
\addlinespace[0.25em]
\Vmol\,$\left(\frac{\rm cm^3}{\rm mol}\right)$ & 12.475 & 12.470\\
\addlinespace[0.25em]
energy gap (eV)                    & 2.2                          & 3.02 \\ 
\addlinespace[0.25em]
\TD\ (K)                           & 1270                         & 1200\\ 
\addlinespace[0.25em]
\bottomrule
\end{tabular}
\label{SiCbasic}
\end{ruledtabular}
\end{table}
The variable $m$ gives the number of Si\,--\,C bilayers consisting of a C and a Si layer stacking in the unit cell. However, most of the available studies refer to the following  polytypes:\cite{SiCinfo} cubic 3C- (zincblende structure, space group F$\bar{4}$3m (T$_{\rm d}^2$); ''ordered'' diamond) and hexagonal 2H-, 4H-, and 6H-SiC (wurtzite, moissanite-4H, and -6H structure, all space group P6$_3$mc (C$^{4}_{6v}$)). The 3C- (2H-) polytype is the only ''pure'' cubic (hexagonal) modification, all other mH-SiC polytypes consist of hexagonal and cubic bonds.\cite{pensl93a} The cubic 3C-modification is also labeled as $\beta$-SiC, whereas the hexagonal polytypes are generally denoted as $\alpha$-SiC. Fig.~\ref{structureSiC} gives a sketch of (a) the diamond-related modification 3C-SiC and (b) the hexagonal 6H-SiC. The C\,--\,Si bilayers are emphasized. In 3C-SiC both elements form face-centered cubic sublattices which are shifted by (1/4, 1/4, 1/4) with respect to each other. Along the $\left<111\right>$ direction the bilayer stacking in 3C-SiC is ABC\,--\dots For the polytypes 2H-, 4H-, and 6H-SiC it is along the $\left<001\right>$ direction ABAB\,--\dots, ABAC\,--\dots, and ABCACB\,--\dots\ Some basic parameters of undoped 3C- and 6H-SiC at room temperature are summarized in Table\,\ref{SiCbasic}.

Depending on the crystal modification, pure SiC exhibits an indirect energy gap between $\sim 2$\,eV (3C-SiC) and $\sim 3.3$\,eV (2H-SiC).\cite{SiCinfo} Slightly doped SiC with donors and acceptors was intensely studied for nitrogen, phosphorus, boron, aluminum, etc.\ by ion-implantation or thermo-diffusion doping. Compared with other dopants, boron was found to have a much faster diffusion rate in SiC. Diffusion processes mediated by the silicon interstitials and by carbon vacancies have been proposed to explain such fast diffusion rates.\cite{bracht00a,rurali02a,gao03a,gao04a} Under silicon-rich conditions the carbon-site substitution is dominating.\cite{bockstedte04a} Among other dopants, the insulator-to-metal transition was observed recently in nitrogen-doped 4H-SiC at carrier concentrations above $10^{19}$\,cm$^{-3}$.\cite{dasilva06a}

In this paper we report a specific-heat study on SiC:B and give a detailed analysis. Moreover, the density-of-states, band dispersions, and two- and three-dimensional plots of the Fermi surfaces for 3C-SiC are presented. We estimate the basic superconducting parameters and compare our findings with the reported results for C:B and Si:B. Finally, we comment on possible origins of the differences between the three superconducting systems.

\section{Experiment}
The preparation and characterization of our samples is described in detail in Ref.~\onlinecite{ren07a}. We studied several samples from different growth processes reproducing the general findings presented in this paper. The particular sample used in this study is identical to that used to map out the \HT\ phase diagram in our previous study, namely ''sample 1'', referred to as ''SiC-1'' in this paper. The hole-doping charge-carrier concentration of SiC-1 was estimated to be $n=1.91\cdot 10^{21}$\,cm$^{-3}$ by a Hall-effect measurement.\cite{nVmolcomment} We note, that all of our so-far prepared samples are polycrystalline materials and not single phase. We found phase fractions of 3C-SiC, 6H-SiC, and Si. In spite of this result the residual resistivity for specimen SiC-1 is already as low as 60\,\mOc. The residual-resistivity ratio ${\rm RRR}=\rho_{\rm 300\,K}/\rho_{\rm 1.5\,K}$ amounts to 10. SiC-1 undergoes a sharp superconducting transition around 1.45\,K. The thermodynamic critical field is estimated to be $\sim 115$\,Oe. In contrast to the type-II superconductivity in C:B and Si:B the nature of the superconductivity in SiC:B is type-I: We find a clear hysteresis in the temperature (field) dependence of the AC susceptibility between cooling (field-down sweep) and subsequent warming (field-up sweep) runs indicating the hallmark of type-I superconductivity as discussed in Ref.\,\onlinecite{ren07a}. 

Specific-heat data was taken by a relaxation-time method using a commercial system (Quantum Design, PPMS). First, we applied a degaussing procedure before the measurement in order to reduce the remanent field of the magnet. Second, the addendum heat capacity was measured at 0\,Oe. Next, specific-heat data was taken in $H=0$\,Oe and subsequently in 200\,Oe, for which the addenda data was not measured, because the difference is expected to be negligibly small. However, this procedure lead to a small but visible artifact in the in-field normal-state data for $0.45\,{\rm K}\leq T \leq 0.6$\,K. Thus, the corresponding data points were removed and not used for the analysis.

\section{Results and Analysis}
\subsection{Specific heat}
\begin{figure}
\centering
\includegraphics[width=8.5cm,clip]{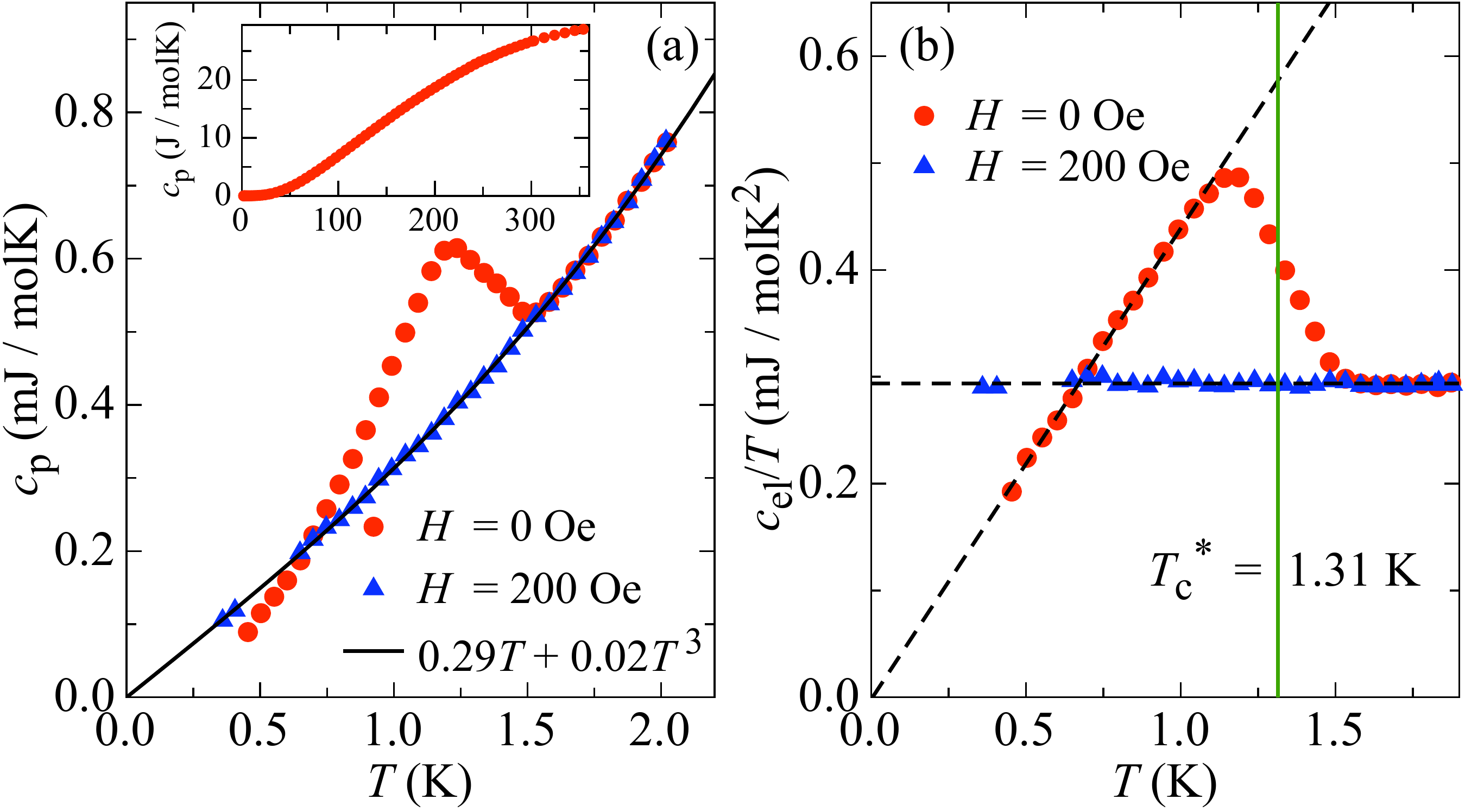}
\caption[]{(color online) Specific heat of SiC-1: The (red) {$\medbullet$} symbols denote the data in zero magnetic field. The (blue) $\blacktriangle$ refer to data measured in a magnetic field $H=200\,{\rm Oe}>\Hc$, representing the normal-state specific heat. (a) Specific heat \cp\ as measured: The line is a fit to the in-field data using the standard Debye formula (Eq.~\ref{GlDebye}). The inset shows the specific heat up to 360\,K. (b) Electronic specific heat $\cel/T$: The lines are an entropy-conserving construction in order to estimate the intrinsic jump height; see text.} \label{cp_SiC-1}
\end{figure}
Fig.~\ref{cp_SiC-1}\,(a) summarizes the temperature dependence of the specific heat \cp\ for specimen SiC-1 in the superconducting state ($H=0$\,Oe) and in the normal-conducting state (achieved by applying a magnetic field $H=200\,{\rm Oe}> \Hc$). The raw data of this figure is the same as that used for Fig.~4 in Ref.~\onlinecite{ren07a} except the removed data points which were affected by an experimental artifact. SiC:B is a bulk superconductor as indicated by the clear jump of \cp\ at \Tc. The inset of Fig.~\ref{cp_SiC-1}\,(a) shows the specific heat up to $\sim 360$\,K for comparison. The room temperature value is still clearly below the classical high-$T$ Dulong-Petit limit (49.88\,J/molK).

The solid curve in Fig.~\ref{cp_SiC-1}\,(a) is a fit to the in-field data for $0.6\,{\rm K} < T < 2$\,K applying the conventional Debye formula
\begin{equation}\label{GlDebye}
\cp=\cph+\cel = \gn T+\beta T^3
\end{equation}
with the Sommerfeld coefficient of the normal-state specific heat \gn\ and the coefficient of the phononic contribution $\beta$ as adjustable parameters. The fit yields a very good description of the data below 2\,K. The obtained values are $\gn=0.29$\,mJ/molK$^2$ and $\beta = 0.02$\,mJ/molK$^4$. From the latter value we determined the Debye temperature \TD\ using $\beta = (12/5)\,\pi^4N \NA\kB/\TD^3$ with the number of atoms per formula unit $N=2$, the Avogadro number \NA, and Boltzmann's constant \kB, yielding $\TD=590$\,K. This is surprisingly low, only half of the value reported for undoped SiC (cf.\,Table\,\ref{SiCbasic}): $\TD^{\rm SiC}\approx 1200\,{\rm K}-1300$\,K. We note that in the case of C:B a significant reduction of the Debye temperature to about 75\,\% of the pure diamond value ($\TD^{\rm C}\approx 1860$\,K, $\TD^{\rm Si}\approx 625$\,K)\cite{ashcroft76} is reported, too.\cite{sidorov05a} In SiC:B this effect turns out to be even more pronounced. 

To further analyze the data we plot the electronic specific heat $\cel=\cp-\cph$ in Fig.~\ref{cp_SiC-1}\,(b). The specific-heat jump starts slightly below $\Tc\approx 1.5$\,K coinciding with the results obtained by our AC susceptibility and resistivity measurements. However, the superconducting transition in \cp\ is rather broad. The lines in Fig.~\ref{cp_SiC-1}\,(b) are an entropy-conserving construction in order to estimate the intrinsic jump hight \Dcel. The ''jump'' temperature $\Tc^*= 1.31$\,K indicated by the perpendicular (green) line in Fig.~\ref{cp_SiC-1}\,(b) is lower than the onset temperature \Tc\ reflecting the broadness of the transition. The jump height is estimated to $\Dcel/\gn\Tc^* \approx 1$. The obtained value is only two thirds of the weak-coupling BCS expectation, namely 1.43. For C:B and other semicondcutor-based superconductors, e.\,g.\ Ge$_{0.95}$Te, an even smaller jump height as low as 0.5 is reported.\cite{sidorov05a,finegold64a} However, the overall shape of the C:B specific-heat data given in Fig.~3 of Ref.~\onlinecite{sidorov05a} is qualitatively different compared to our data. The authors report a very broad transition consisting of two well-separated transitions.

Next we focus on the question of the superconducting gap symmetry. We try the following two models to describe our experimental data: 

Model (i) \textit{Assuming an isotropic gap structure:}

The simplest approach to obtain information about the superconducting gap is given by the conventional BCS text-book formula\cite{tinkham96} $\cel(T)/T\propto \exp(-\Delta(0)/T)/T$. However, paying respect to the facts that on one hand the exponential behavior is only expected well below \Tc\ and on the other hand data below approximately 0.45\,K is lacking leads to the idea to replace the exponential term by tabulated numerical specific-heat data calculated in the standard weak-coupling BCS framework,\cite{muehlschlegel59a} which is in principal valid up to \Tc. Therefore, we fitted the tabulated data with a polynomial $\cel^{\rm BCS}$ (15$^{\rm th}$ order) leading to
\begin{equation}\label{GlBCS_neu}
\cel(T)/T \propto \cel^{\rm BCS}(T)/T.
\end{equation}
Next, considering that the samples are not single phase, it is reasonable to assume an additional $T$-linear term $\gres T$ reflecting a residual density of states originating from non-superconducting metallic inclusions. This modifies Eq.\,\ref{GlBCS_neu} as follows:
\begin{equation}\label{GlBCS_res}
\cel(T)/T = \gres + \gs\cdot \cel^{\rm BCS}(T)/T.
\end{equation}
Since the entropy related to a residual term $\gres\Tc^*$ does not contribute to the specific-heat jump, the prefactor of the BCS term is given by $\gs=\gn-\gres$. Therefore \gres\ is the only adjustable parameter in this approach.

Model (ii) \textit{Assuming a power-law behavior of the electronic specific heat:}

At temperatures well below \Tc, a superconducting gap structure with nodes leads to the power-law behavior
\begin{equation}\label{Glpower}
\cel(T)/T = \gres + a\cdot T^b
\end{equation}
with $b=1$ or 2 for line or point nodes.\cite{volovik85a,tinkham96} Here, we pay respect to a residual contribution, too.

\begin{figure}
\centering
\includegraphics[width=7.5cm,clip]{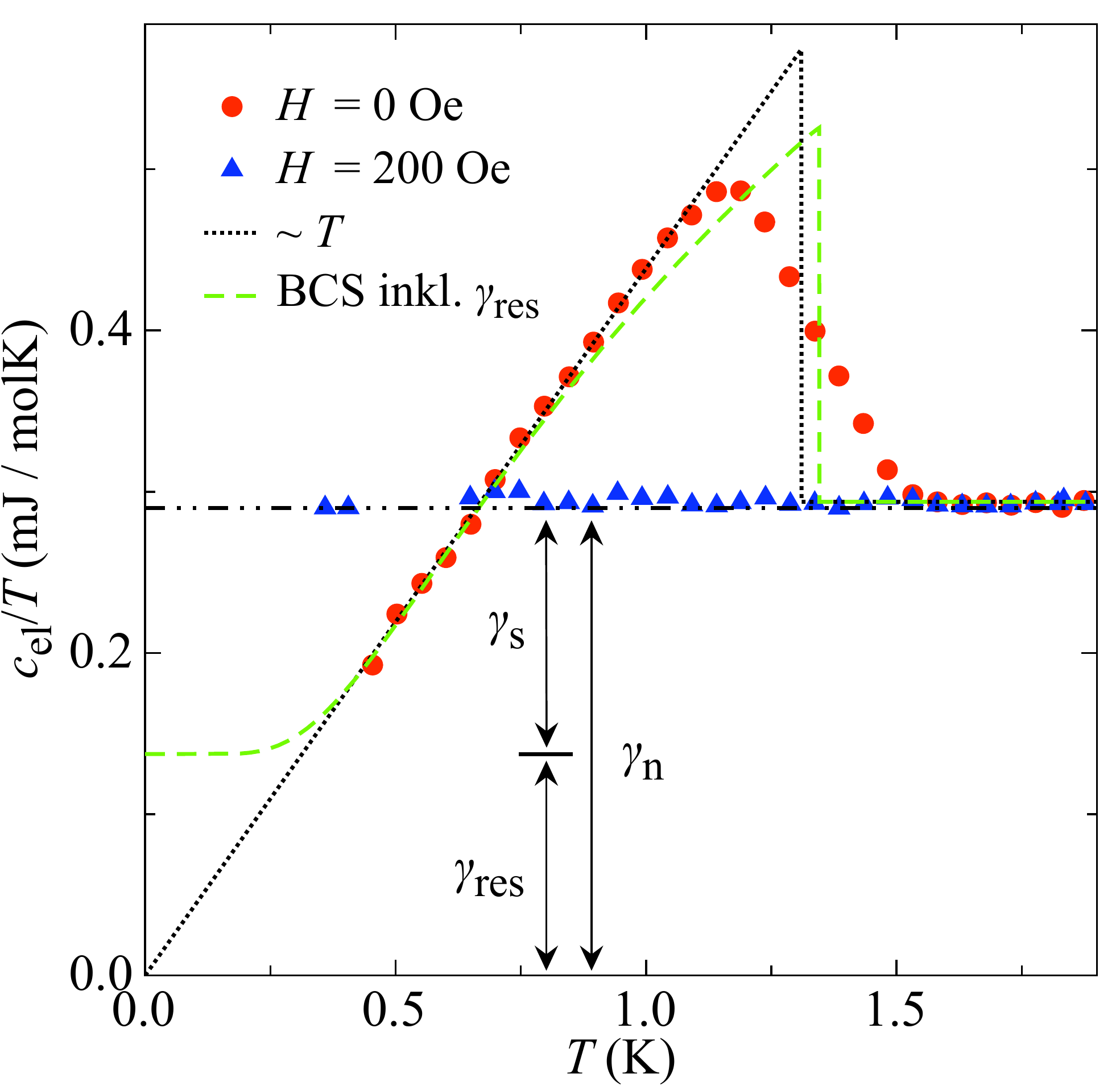}
\caption[]{(color online) Electronic specific heat of SiC-1: The (red) {$\medbullet$} / blue $\blacktriangle$ symbols denote again zero-field / in-field data. The lines are fits to the data assuming an isotropic gap (dotted line) and a nodal gap (dahsed line). The dashed-dotted horizontal line denotes the value of the Sommerfeld parameter in the normal-conducting state \gn. The two other $\gamma$ values are estimates related to model (i) and denote the superconducting \gs\ and a possible residual Sommerfeld parameter \gres; see text.} \label{cel_SiC-1_sc_fits}
\end{figure}
The results obtained by applying both models are summarized in Fig.~\ref{cel_SiC-1_sc_fits}. Applying Eq.\,(\ref{GlBCS_res}) to the data yields the green dashed curve, Eq.\,(\ref{Glpower}) the black dotted line. The estimated $\gamma$ factors for model (i) are given, too: \gn\ (normal-conducting state), \gs\ (superconducting state), \gres\ (residual contribution).

\textbf{Model (i)}
Applying Eq.\,(\ref{GlBCS_res}) to the data for $T<0.7$\,K gives a reasonable description as seen in Fig.~\ref{cel_SiC-1_sc_fits} (green dashed line). The residual $\gamma$ coefficient amounts to $\gres=0.14$\,mJ/molK$^2$. Above 0.7\,K the fit undershoots the data corresponding to a distribution of \Tc\ values reflected in the broadness of the transition. An entropy-conserving construction using the fit result instead of the linear approximation shown in Fig.~\ref{cp_SiC-1}\,(b) yields a slightly higher ''jump'' temperature $\Tc^*=1.35$\,K due to the downwards curvature around \Tc. Paying respect to the residual contribution $\gres\Tc^*$ and evaluating the jump height with the Sommerfeld parameter of the superconducting part of the sample, $\gs=0.16$\,mJ/molK$^2$, gives almost the value predicted by the BCS theory: $\Dcel/\gs\Tc^*=1.48$.

\textbf{Model (ii)} 
The assumption of a power-law behavior in order to describe the data yields an even better description: Assuming a linear temperature dependence of $\cel/T$ reproduces the experimental data in the whole temperature range below $\approx 1.1$\,K, i.\,e.\ below the transition down to 0.45\,K and extrapolates further down to 0\,K without any indication of a residual contribution $\gres\Tc^*$ in contrast to the results obtained by applying model (i) to the data. The resulting fitting curve is shown in Fig.~\ref{cel_SiC-1_sc_fits} (dotted black line). It was obtained by adjusting only the prefactor $a$ and fixing $b=1$ and $\gres=0$ in Eq.\,(\ref{Glpower}). The attempt to include a residual term to the fit gave $\gres\approx 0$. A $T$-linear behavior of $\cel/T$ is expected in the case of a gap containing line nodes, but only well below \Tc\ where the superconducting gap is nearly independent of temperature. It is expected that at higher temperatures $T\rightarrow \Tc$ the specific heat is affected by the reduction of the gap magnitude and therefore deviates from the linear extrapolation.\cite{hasselbach93a} A complex balance of different effects is needed to cause an apparent linear temperature dependence up to \Tc. We note, that a $T$-linear behavior of $\cel/T$ up to \Tc\ has been reported for e.\,g.\ the heavy-fermion compounds URu$_2$Si$_2$ and UPt$_3$.\cite{hasselbach93a,fisher89a}

The obtained jump height using a linear entropy-conserving construction, $\Dcel/\gn\Tc^* \approx 1$ (Fig.~\ref{cp_SiC-1}\,(b)), is similar to the value predicted theoretically for a superconductor with a nodal gap structure.\cite{hasselbach93a,nishizaki99a} We note that in such a model $\cel/T$ should exhibit a rounded maximum at \Tc\ rather than the triangle-like peak used in the simplified entropy-conserving construction shown in Fig.~\ref{cp_SiC-1}\,(b). Therefore the jump height would be even smaller than 1 and $\Tc^*$ slightly higher.

However, in spite of the satisfying description of the data following model (ii) it is still necessary to obtain data down to several 10\,mK to clarify the true nature of the superconducting gap. It would not be surprising if the specific heat of the multi-phase sample used consists of an additional $T$-linear term due to a residual \gres\ as suggested by the result of model (i). Therefore the apparent power-law behavior of the experimental electronic specific heat for $0.45\,{\rm K} < T < 1.1$\,K extrapolating to 0 for $T\rightarrow 0$\,K is rather striking.

\subsection{Superconducting Parameters}
Together with the resistivity, Hall-effect, and AC susceptibility data published in Ref.~\onlinecite{ren07a} we are able to estimate the basic superconducting parameters. They are summarized in Table~\ref{SiCprop} along with the derived normal-state parameters. For comparison the so-far known corresponding parameters for C:B and Si:B are listed, too.
\begin{table}[t]
\centering
\begin{ruledtabular}
\caption{Normal-state and superconducting properties of SiC:B compared to those reported for C:B (Refs.~\onlinecite{ekimov04a} and \onlinecite{sidorov05a}) and Si:B (Ref.~\onlinecite{bustarret06a}). Note that the highest \Tc\ (\Hcz) for C:B reported so far is 11.4\,K ($8.7\cdot 10^4$\,Oe).\cite{umezawa05a} The asterisked ''*$\dots$*'' values are preliminary because they depend on the value of $\rho_0$ which we believe is still not the intrinsic one; see text. The coherence length $\xi$ in the case of the type-II superconductor C:B was estimated using Eq.\,(\ref{cohlength1}) (i.\,e.\ from $n$, \gn, and \Tc) for better comparison with the type-I superconductor SiC:B. The numbers given in parantheses are the published values from Refs.~\onlinecite{bustarret06a} (Si:B) and \onlinecite{sidorov05a} (C:B) calculated with Eq.\,(\ref{cohlength2}) (i.\,e.\ from \Hcz).}
\label{SiCprop}
\begin{tabular}{lccc}
\toprule
                      & SiC:B  & C:B    & Si:B \\ \hline \addlinespace[0.75em]
$n$ (cm$^{-3}$)       & $1.91\cdot 10^{21}$ & $1.80\cdot 10^{21}$ & $2.80\cdot 10^{21}$ \\ \addlinespace[0.1em]
$\gn$ (mJ/molK$^2$)   & 0.294  & 0.113  & --   \\ \addlinespace[0.1em]
$\beta$ (mJ/molK$^4$) & 0.0193 & 0.0007 & --   \\ \addlinespace[0.1em]
\TD\ (K)              & 590    & 1440   & --   \\ \addlinespace[0.1em]
$\Dcel/\gn\Tc^*$      & 1      & 0.50   & --   \\ \addlinespace[0.1em]
$\rho_0$ (\mOc)       & *60*   & 2500   & 130  \\ \addlinespace[0.1em]
RRR                   & *10.0* & 0.9    & 1.2  \\ \addlinespace[0.1em]
$\Tc(0)$ (K) (onset)  & 1.45   & 4.50   & 0.35 \\ \addlinespace[0.1em]
$\Hc(0)$ (Oe)         & 115    & --     & --   \\ \addlinespace[0.1em]
$\Hsc(0)$ (Oe)        & 80     & --     & --   \\ \addlinespace[0.1em]
$\Hcz(0)$ (Oe)        & type-I & $4.2\cdot10^4$ & 4000   \\ \addlinespace[0.1em]
\kF\ (nm$^{-1}$)      & 3.8    & 3.8    & --   \\ \addlinespace[0.1em]
$m^*$ (\me)           & 1.2    & 1.7    & --   \\ \addlinespace[0.1em]
$\tau(0)$ (fs)        & *37*   & --     & --   \\ \addlinespace[0.1em]
\vF\ (m/s)            & $3.8\cdot 10^5$ & -- & --   \\ \addlinespace[0.1em]
$\ell$ (nm)           & *14*   & 0.34   & --   \\ \addlinespace[0.1em]
$\xi(0)$ (nm)         & 360    & 80 (9) & (20) \\ \addlinespace[0.1em]
$\lambda(0)$ (nm)     & 130    & 160    & --   \\ \addlinespace[0.1em]
\kGL(0)               & 0.35   & 2 (18) & --   \\ \addlinespace[0.1em]
\bottomrule
\end{tabular}
\end{ruledtabular}
\end{table}
From the charge-carrier concentration\cite{nVmolcomment} ($n=1.91\cdot 10^{21}$\,cm$^{-3}$) assuming a single spherical Fermi surface we obtain the Fermi-wave number $\kF = (3\pi^ 2n)^{1/3}=3.8$\,nm$^{-1}$. The effective mass is evaluated as $m^*=(3\hbar^2\gn)/(\Vmol \kB^2 \kF)=1.2\me$ with the bare-electron mass \me\ and the molar volume\cite{nVmolcomment} \Vmol. The Fermi velocity $\vF=\hbar\kF/m^*$ amounts to about 0.1\,\% of the speed of light. The mean-free-path is estimated to be $\ell = \hbar\kF/(\rho_0 n e^2)=14$\,nm with the elementary charge $e$. The superconducting penetration depth amounts to $\lambda(0)= \sqrt{m^*/(\mu_0 n e^2)}=130$\,nm. The coherence length is estimated using the BCS expression\cite{tinkham96}
\begin{equation}
\xi(0)=0.18\hbar\vF/(\kB\Tc),
\label{cohlength1}
\end{equation}
which yields $\xi(0)=360$\,nm. Hence, the Ginzburg-Landau parameter\cite{kGLcomment} is $\kGL=0.96\lambda(0)/\xi(0)=0.35< 1/\sqrt{2}$, clearly placing SiC:B in the type-I regime. 

\begin{figure}
\centering
\includegraphics[width=7.5cm,clip]{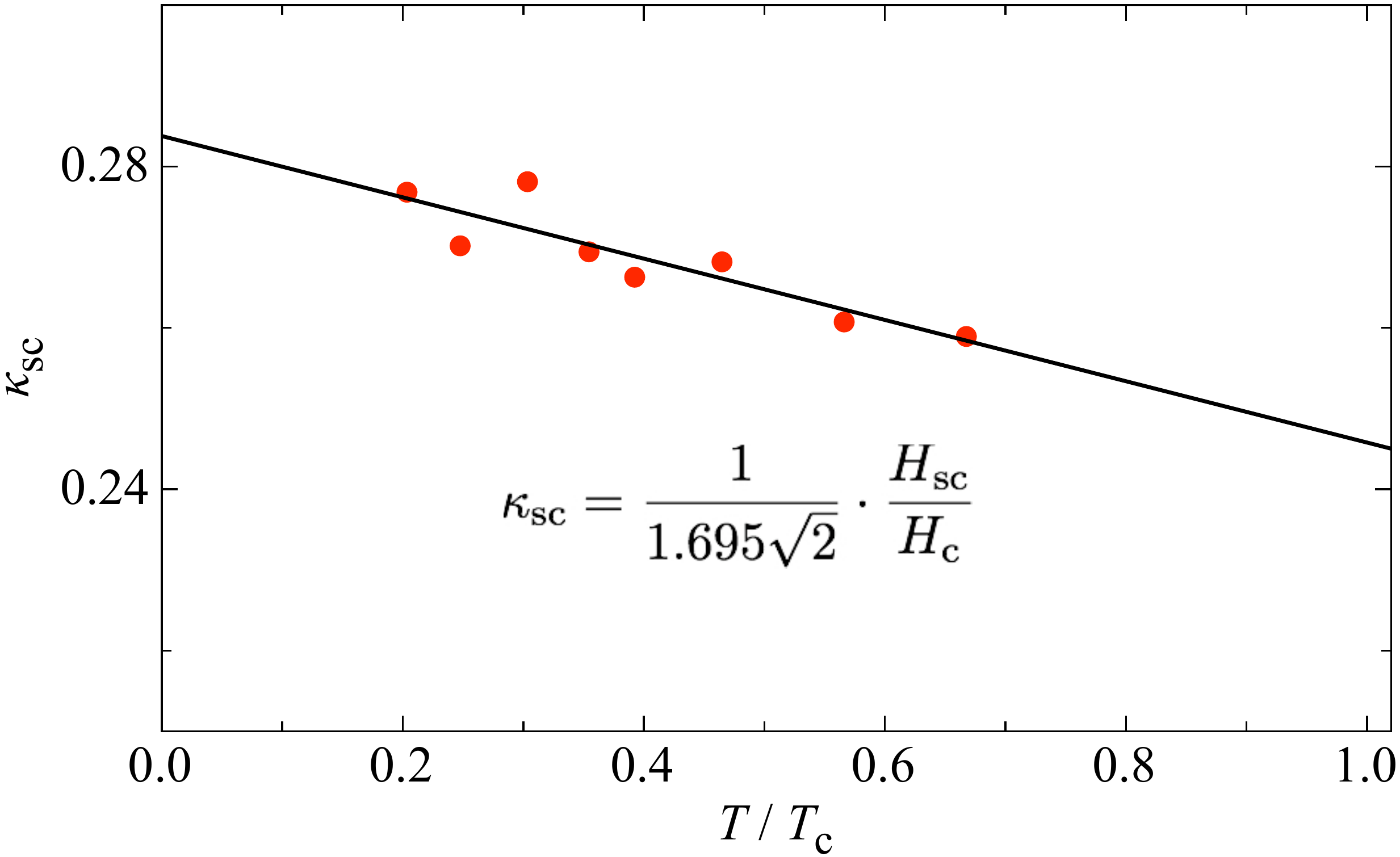}
\caption[]{(color online) Ratio of $\Hc/\Hsc$ according to Eq.\,(\ref{GLsc}): The solid line is a linear fit to the data; see text.}
\label{HTanalysis}
\end{figure}
An independent approach to estimate the GL parameter \kGL\ is to start with the two phase lines given in the \HT\ phase diagram (Fig.\,5 in Ref.\,\onlinecite{ren07a}), i.\,e.\ the supercooling field \Hsc\ and the critical field strength \Hc. Following the GL theory the analysis of the observed supercooling behavior in AC susceptibility provides an upper limit of the GL parameter. Upon decreasing an external applied magnetic field at constant temperature the superconducting {\it nucleation field} is given by \Hcz. For a type-I superconductor this is smaller than the thermodynamic {\it critical field} \Hc, which leads to the effect of {\it supercooling} if moreover the GL parameter satisfies $\kGL<0.417$. However, Saint-James and de~Gennes showed that the nucleation of superconducting parts of a sample sets in near to the surface at the {\it surface nucleation field} $\Hcd = 1.695\Hcz = 1.695 \sqrt{2}\kGL\Hc$, larger than \Hcz\ if the sample is placed in vacuum.\cite{saintjames63a,Hc3comment} In real experiments, however, the onset of superconductivity will be observed at a field \Hsc, which is larger than the ideal supercooling \Hcd: the experimentally derived supercooling phase line in an \HT\ phase diagram will satisfy the inequality $\Hsc \geq \Hcd$. Hence, the ''real'' GL parameter has to be smaller than the ''supercooling'' $\kappa_{\rm sc}$:\cite{feder69a,tinkham96}
\begin{equation}\label{GLsc}
\kGL\leq \kappa_{\rm sc}= \frac{1}{1.695\sqrt{2}}\cdot\frac{\Hsc}{\Hc}=0.417\frac{\Hsc}{\Hc}.
\end{equation}
We would like to emphasize that the qualitative observation of supercooling in the field-dependence of our AC susceptibility data already confirms that the GL parameter has to be {\it smaller} than 0.417.

We calculated the ratio of $\Hsc$ and $\Hc$ deduced from field-sweep measurements at several temperatures. The result is shown in Fig.~\ref{HTanalysis}. Following the procedure used by Feder and McLachlan,\cite{feder69a} i.\,e.\ extrapolating the data to $T=\Tc$, yields in our case $\kGL < 0.3$, which is even smaller than the afore reported estimate derived from $n$, \Tc, and \gn, supporting the conclusion that SiC:B is a type-I superconductor.

However, our results also imply that SiC:B is a dirty-limit superconductor because the coherence length is much larger than the mean-free path: $\xi(0)\gg \ell$. The Ginzburg-Landau parameter for a dirty-limit superconductor\cite{tinkham96} is given by $\tilde{\kappa}_{\rm GL}=0.715\lambda(0)/\ell$ which is $\gg 1/\sqrt{2}$ in the sample used due to the small $\ell$ value. The large $\tilde{\kappa}_{\rm GL}$ / small $\ell$ is mainly caused by the residual resistivity $\rho_0$. Among our samples the residual resistivity varies from 60\,\mOc\ to the order of m$\Omega$cm, all of them exhibiting a type-I behavior in the AC susceptibility. With this experimental finding and keeping in mind that the so-far prepared crystals are polycrystalline multi-phase materials, it is reasonable to assume that the intrinsic value of $\rho_0$ (and hence $\ell$) could be much lower (larger) than even the 60\,\mOc\ (14\,nm) found for the sample SiC-1. The quantities given in Table~\ref{SiCprop} which are related to the value of $\rho_0$ are not very reliable and therefore asterisked ''*$\dots$*''. A decrease of the residual resistivity to a few \mOc\ would be sufficient to shift $\tilde{\kappa}_{\rm GL}$ below the critical value of $0.417<1/\sqrt{2}$ in accordance with our experimental finding of a supercooled type-I superconductor.

\subsection{Band structure of 3C-SiC}
\begin{figure}
\centering
\includegraphics[width=8.5cm,clip]{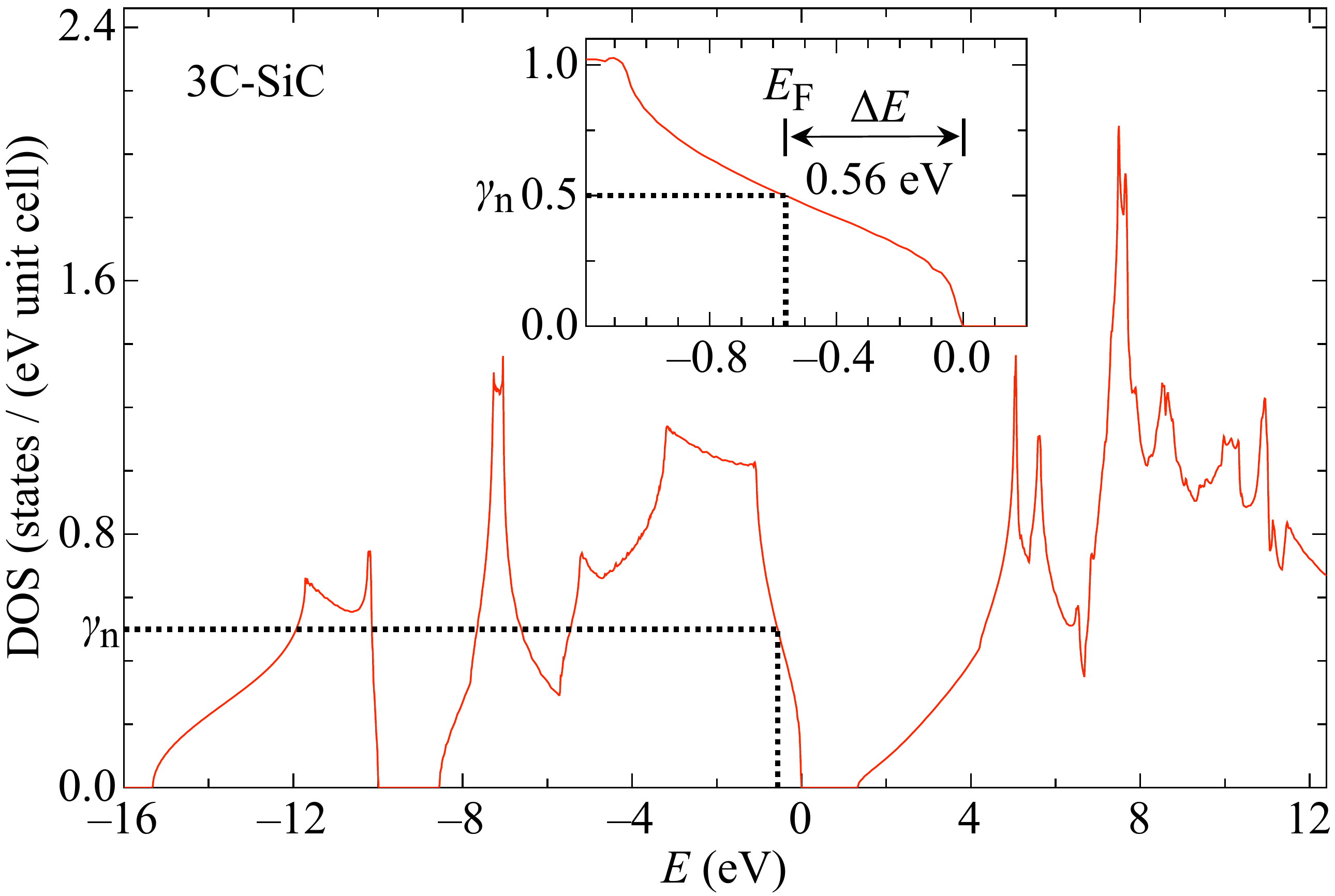}
\caption[]{(color online) Calculated density of states (DOS) vs. energy for zincblende 3C-SiC. The inset gives an enlarged view of the energy range near to the Fermi level \EF. The origin of energy is taken at the valence-band maximum without doping. The dotted lines in both panels mark the experimental value of the Sommerfeld parameter $\gn=0.29$\,J/molK$^2$\,$=0.5$\,/(eV unit cell) and the respective energy shift due to the boron doping $\Delta E=0.56$\,eV assuming rigid bands; see text for details.}
\label{3C-SiC_DOS}
\end{figure}
Calculated density of states (DOS) and band structure data provide another possibility to determine an upper limit of the Fermi-wave number and hence the GL parameter \kGL\ using the experimental value of the Sommerfeld coefficient and 
\begin{equation}
\gn = \frac{\pi^2\kB^2}{3}\cdot \sum_{i=1}^{3}{\rm DOS}_{i}(E_{\rm F})=0.29\,{\rm mJ/molK}^2.
\label{DOS}
\end{equation}

For simplicity, we will focus only on the 3C-modification of SiC. We approximate, that all three valence bands are free-electron like. Moreover, we assume rigid bands, i.\,e.\ the band structure is independent of charge-carrier doping. 

\begin{figure}
\centering
\includegraphics[width=8.5cm,clip]{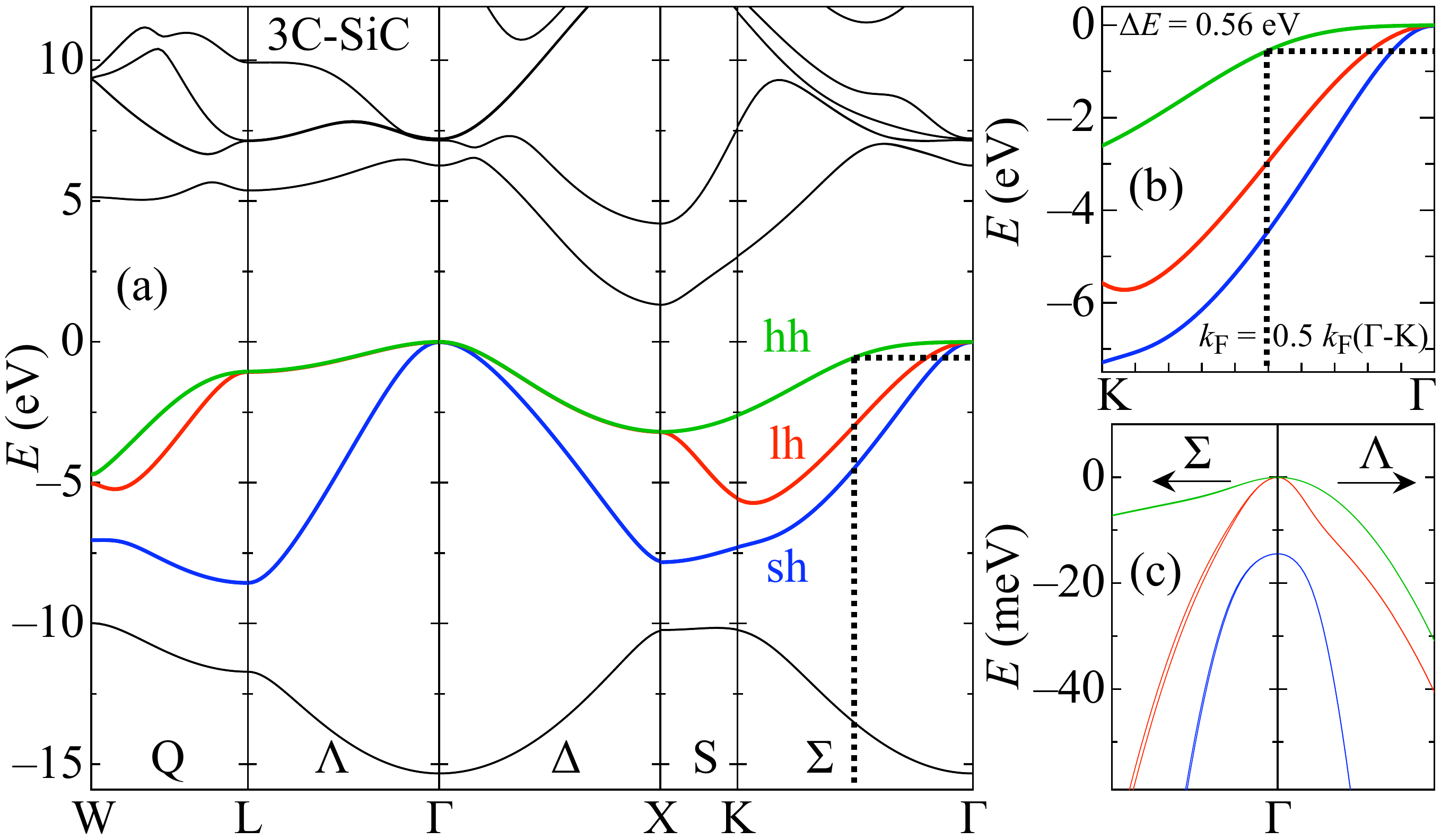}
\caption[]{(color online) Band-structure calculation of 3C-SiC. Spin-orbit coupling is included. The upper three valence bands are the heavy-hole (hh, green), the light-hole (lh, red), and the split-hole (sh, blue) band. Panel (a) summarizes the band-structure calculation based on the DOS calculation given in Fig.\,\ref{3C-SiC_DOS}. Panel (b) gives an enlarged view along the [110] direction from the Brillouin zone center $\Gamma$ towards the K point. The dotted lines mark the energy shift $\Delta E=0.56$\,eV estimated from the data shown in Fig.\,\ref{3C-SiC_DOS} and the corresponding Fermi-wave number $\kF=0.5\cdot\kF(\Gamma-{\rm K})= 0.5\cdot \sqrt{18}/4\cdot 2\pi/a$; see text for details. Panel (c) gives an enlarged view of the zone center at $\Gamma$. The bands are split due to the spin-orbit coupling.}
\label{3C-SiC_BS}
\end{figure}

\begin{figure}
\centering
\includegraphics[width=8.5cm,clip]{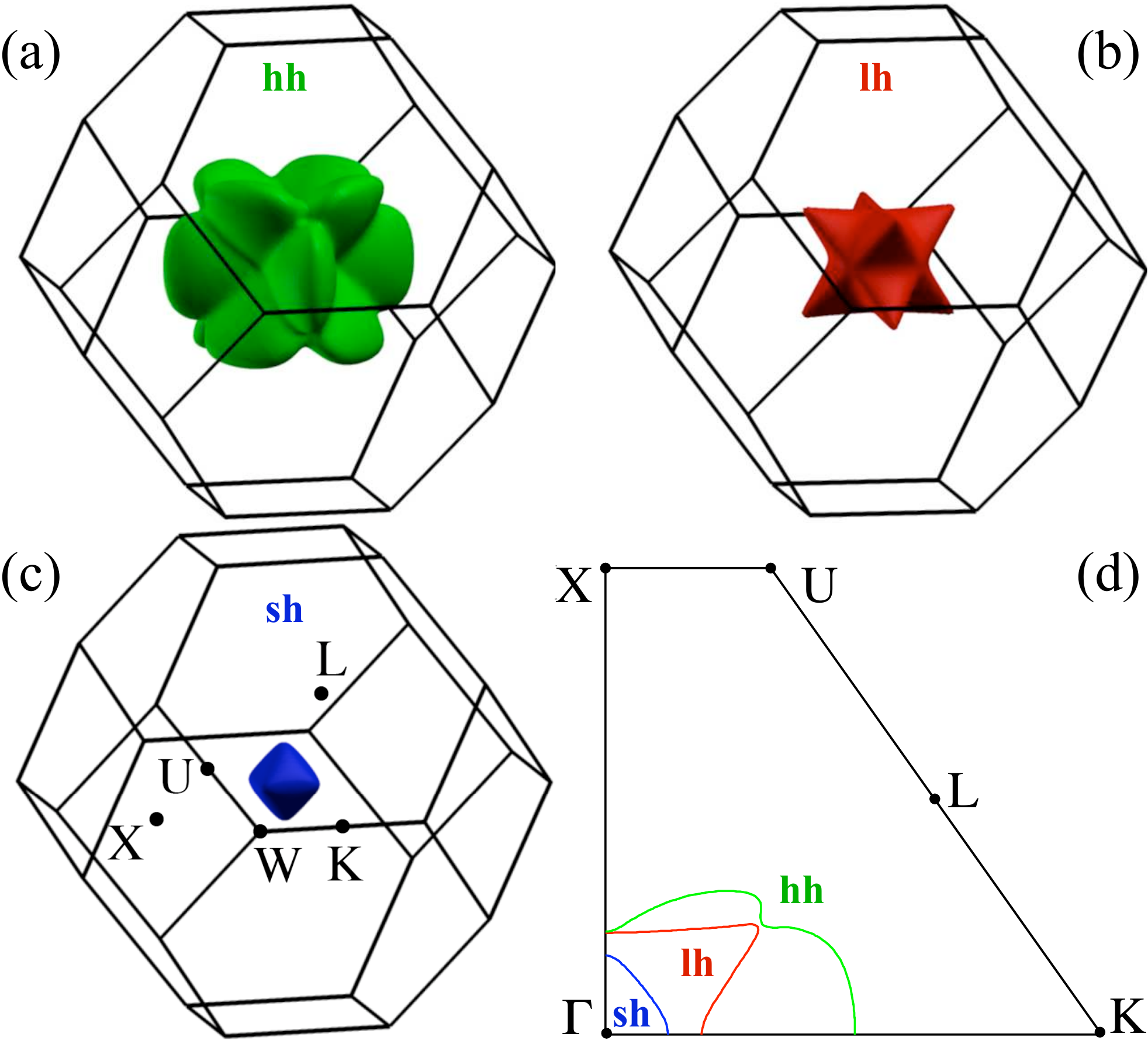}
\caption[]{(color online) Plots of the Fermi 
surfaces of 3C-SiC around the $\Gamma$ point of the fcc Brillouin zone (given in black) for a Fermi energy below the top of the valence band by $-0.56$\,eV. Only three instead of six bands are shown due to the smallness of the spin-orbit splitting. Panels (a), (b), and (c) display three-dimensional representations of the heavy-hole (hh), the light-hole (lh), and the split-hole (sh) band. Panel (d) contains their cross sections. The colors of the respective band plots are the same as used in Fig.\,\ref{3C-SiC_BS}.}
\label{3C-SiC_FS}
\end{figure}

The electronic band structure of 3C-SiC is calculated within the local density approximation (LDA) to the density functional theory. The all-electron full-potential linear-augmented-plane-wave method is used to solve one-electron Kohn-Sham equations. All the relativistic effects including spin-orbit coupling are included to every self-consistent-field iteration. The results for the total DOS and the band dispersions are shown in Figs.\,\ref{3C-SiC_DOS} and \ref{3C-SiC_BS}, respectively. Two- and three-dimensional plots of the three Fermi surfaces corresponding to the upper three valence bands of 3C-SiC are given in Fig.\,\ref{3C-SiC_FS}. Therein panel (a), (b), and (c) display three-dimensional representations of the heavy-hole (hh), the light-hole (lh), and the split-hole (sh) bands. Panel (d) gives their cross sections.\cite{persson97a}

The DOS corresponding to the experimental \gn\ value is indicated by a dotted line in Fig.\,\ref{3C-SiC_DOS}. The inset gives an expanded view of the relevant energy range. Using Eq.\,(\ref{DOS}), the molar volume \Vmol\ and the volume of the unit cell $V_0$ (cf.\ Table\,\ref{SiCbasic}) to convert the corresponding units, $\gn =0.29$\,mJ/molK corresponds to ${\rm DOS}\approx 6\cdot 10^{21}$\,states\,/\,(eV$\cdot$cm$^3$) = 0.5\,states\,/\,(eV$\cdot$ unit cell). The energy shift of the Fermi energy due to the charge-carrier doping was estimated to $\Delta E= 0.56$\,eV also indicated by a dotted line in Fig.\,\ref{3C-SiC_DOS}.

The band structure of 3C-SiC is shown in Fig.\,\ref{3C-SiC_BS}\,(a). Panel (b) provides an enlarged view of the relevant bands near the $\Gamma$ point. In panel (c) the effect of spin-orbit coupling at the $\Gamma$ point is shown, which is about 10\,meV.

Next we plot the estimated energy shift $\Delta E$ into the band structure plot as marked by a dotted line in Figs.\,\ref{3C-SiC_BS}\,(a) and (b). The corresponding Fermi-wave number for the heavy-hole band was estimated to $\kF\approx 50$\,\% of the distance from $\Gamma$ to K in the fcc Brillouin zone which equals $\sqrt{18}/4\cdot 2\pi/a$. The parameter $a$ denotes the lattice constant of 3C-SiC $a=4.3596$\,\AA\ (cf.\ Table\,\ref{SiCbasic}). This yields an upper limit of $\kF< 7.6$\,nm$^{-1}$, which is more than double the value evaluated before (3.8\,nm$^{-1}$) assuming a single spherical Fermi surface. 

Possible origins of the discrepancy between the two estimates of \kF\ could be the neglect of the 6H-SiC phase fraction. Moreover, the assumption of a single spherical Fermi surface is too simple as can be seen in Fig\,\ref{3C-SiC_FS}. It is not known how the heavy-boron doping modifies the real DOS and band structure, either, leading back to the question if the superconductivity in this material evolves from intrinsic or impurity bands as discussed in literature for C:B. 

\section{Comparison and Discussion}
After the analysis of the experimental data we now focus on the question why the ''mixed'' compound SiC is a type-I superconductor whereas the ''pure'' parent compounds silicon and diamond exhibit type-II superconductivity upon boron doping. Let us therefore briefly compare the obtained parameters of SiC:B with those reported for C:B:\cite{sidorov05a} For Si:B no specific-heat study is available so far, hence a comparison is not possible for all parameters. 

The charge-carrier concentrations for all three specimen are comparable and on the order of $2\cdot 10^{21}$\,cm$^{-3}$. However, the temperature dependence and the absolute values of the resistivity are different: SiC:B turns out to be a much better conductor exhibiting a metallic $\rho(T)$ for $\Tc\leq T \leq 300$\,K with an RRR value of about 10, whereas the resistivity of C:B and Si:B decreases slightly above \Tc. For Si:B the slope of the resistivity becomes positive above $\sim 50$\,K, for C:B a slightly positive slope is observed only above 200\,K.\cite{sidorov05a,bustarret06a} In both cases the resistivity is almost temperature independent resulting in RRR values of about 1.

The Sommerfeld parameter $\gn$ is somewhat smaller for C:B compared to SiC:B, the coefficient of the phononic contribution is much smaller for C:B resulting in a higher Debye temperature in the latter case. This is not surprising since the Debye temperature of pure diamond is much higher than that of pure SiC. For both compounds the jump height of the specific heat (Fig.~\ref{cp_SiC-1}\,(b)) is much smaller than the BCS expectation for a weak-coupling superconductor. The superconducting penetration depths are similar for C:B and SiC:B ($\lambda(0)\approx 150$\,nm) but the coherence lengths make the essential difference. The published values from Refs.\,\onlinecite{bustarret06a} (Si:B) and \onlinecite{sidorov05a} (C:B) have been estimated from the upper critical field strength \Hcz\ using the GL expression
\begin{equation}
\xi = \sqrt{\Phi_0/2\pi\Hcz(0)}
\label{cohlength2}
\end{equation}
with the flux quanta $\Phi_0$. Applying this formula the coherence lengths of C:B and Si:B are both on the order of 10\,nm. These values of $\xi$ are given in parantheses in Table~\ref{SiCprop}. In the case of C:B we calculated $\xi(0)$ using Eq.\,(\ref{cohlength1}) for a better comparability with SiC:B, too. The latter yields $\xi(0)=80$\,nm, whereas for SiC:B $\xi(0)$ amounts to 360\,nm resulting in different GL parameters $\kGL= 0.35$ for SiC:B and 2 for C:B (which is 18 using the published value of $\xi=9$\,nm), cf.\ Table~\ref{SiCprop}. Hence, SiC:B is a type-I and C:B a type-II superconductor.

At the current state of research we can only speculate about the physical reasons for this different nature of superconductivity in C:B / Si:B and SiC:B. In the case of C:B one apparent reason leading to a smaller coherence length and hence a larger GL parameter is the higher critical temperature of this superconductor: $\xi(0)\propto \Tc^{-1}$. However, this argument does not hold for Si:B, \Tc\ of which is much smaller than that of SiC:B. One can argue that SiC:B is a much cleaner system than C:B and Si:C. Hence, the coherence length $\xi(0)$ of Si:B (thin film and diffuse doping) might be limited by a very short mean-free path $\ell$ and thus the GL parameter is larger, leading to the speculation that ''clean'' Si:B could be a type-I superconductor, too.

Finally, we would like to mention a couple of apparent differences between the systems: 

\textbf{Si\,--\,C bilayers:} SiC is in a certain sense a ''layered'' system consisting of Si\,--\,C bilayers. Many polytypes are known distinguished by the stacking sequence of these bilayers in the crystal unit cell. In this sense one may refer to SiC as an ''ordered system''. Our results\cite{ren07a} suggest that boron is introduced only into the carbon sites in SiC:B and hence only half of the crystal sites are directly affected by the disorder due to the hole-doping process, whereas in C:B and Si:B in principle all sites can be randomly involved. 

\textbf{structure:} For silicon and diamond the cubic crystal structure seems to be important or a precondition for the appearance of superconductivity upon doping. In SiC the situation is different. The multi-phase crystal used in this study contains cubic 3C-SiC {\it and} hexagonal 6H-SiC. At the moment we cannot rule out the possibility that {\it both} phase fractions contribute to the superconductivity which would be a clear difference compared to the two parent compounds. Moreover, Cohen suggested in 1964 that {\it hexagonal} SiC could exhibit superconductivity.\cite{cohen64b} However, the same author predicts that most of the many-valley semiconductor based superconductors should be type-II rather than type-I.\cite{cohen64a} 

\textbf{band structure:} In contrast to cubic diamond and silicon the ''mixed'' compound SiC breaks inversion symmetry. In crystals with inversion-symmetry, states with different spin orientations are degenerated. This is not true in general for crystals with broken inversion symmetry. The degeneracy might be lifted by the spin-orbit interaction. In 3C-SiC (zincblende structure) the degeneracy is preserved only along the $\left[100\right]$ direction. Along e.\,g.\ the $\left[110\right]$ direction (i.\,e.\ from the $\Gamma$ point to the K point in the Brillouin zone) the states with different spins split up.\cite{cardona88a,theodorou99a} Using the above estimate of the Fermi-wave number for the heavy-hole band $\kF<7.6$\,nm$^{-1}$ we can give a rough estimate of the spin-orbit splitting for this band along $\Gamma$\,--\,K in 3C-SiC:\cite{theodorou99a} $\Delta_{\rm SO}=0.02$\,meV, which is a rather small value. We note, that the light-hole, split-hole, and (in the case of electron doping) the lowest-conduction band exhibit larger values.\cite{theodorou99a} Using as an example $\kF=3.8$\,nm$^{-1}$ gives the same spin splitting because $\Delta_{\rm SO}$ of the heavy-hole band is almost constant in the interval $0.25\cdot 2\pi/a$ to $0.85\cdot 2\pi/a$ along the [110] direction of the Brillouin zone.\cite{theodorou99a}

\textbf{inversion symmetry:} Moreover, a broken inversion symmetry is known to give rise to a highly interesting nature of the superconducting ground state, including a spin singlet -- triplet mixture.\cite{frigeri04a,fujimoto07a} In SiC:B we do not expect any unconventional scenario based on the broken inversion symmetry because of the comparably light elements silicon and carbon without strong electron-electron interaction. 

\textbf{charge-carrier concentration:}
In diamond, cubic silicon, and 3C-SiC the indirect band gaps are between the zone center ($\Gamma$ point) and the X point of the Brillouin zone. For all other SiC polytypes the valence-band maximum is located at the $\Gamma$ point, too, but the conduction-band minimum differs. For 6H-SiC it occurs at the M point.\cite{pensl93a,park94a,harris95} If one takes into account spin-orbit interaction, the splitting of the band structure of diamond and silicon is not affected, but for 3C-SiC the splitting will cause a shift of the valence-band maximum.\cite{burns77} However, the boron doping in SiC removes electrons from the valence bands and therefore the difference in the semiconducting gaps might be of minor relevance.

Nevertheless, it underlines again the importance of answering the question whether the holes induced by boron doping in SiC reside in the intrinsic bands or form an impurity band, i.\,e.\ what is the nature of the metallic ground state, from which superconductivity develops? 

\section{Summary}
In summary we present a specific-heat study of heavily boron-doped silicon carbide SiC:B using the same crystal used in our recent publication reporting the discovery of superconductivity. In contrast to the type-II superconductivity in the two parent compounds, boron-doped diamond C:B and boron-doped silicon Si:B, the bulk superconductivity in SiC:B is type I. This is reflected in rather different values of the superconducting coherence length, i.\,e.\ 360\,nm for SiC:B and only 80\,nm for C:B, whereas the penetration depths are of the same order of magnitude. We presented two different approaches to describe the data: assuming (i) an isotropic gap structure and (ii) a power-law behavior. The electronic specific heat in the superconducting state is well reproduced by the former assumption with a residual density of states or by the latter assumption of a quadratic temperature dependence. The specific-heat jump height $\Delta \cel/T$ at \Tc\ is about 1 or even smaller in the latter model and hence far away from the expectation in a BCS framework. However, due to the lack of data points below 0.45\,K it is difficult to give a final conclusion about the superconducting gap structure. To further clarify the gap structure a specific-heat study in a dilution refrigerator system is desired.

The origin of the different nature between the type-II superconductors C:B and Si:B on the one hand and the type-I superconductor SiC:B on the other hand remains at this state of research unclear. To clarify this intriguing issue further experimental and theoretical work is needed. From the experimental point of view single crystalline samples are highly desirable. Moreover samples with only one phase fraction, either 3C-SiC or 6H-SiC, are eligible to answer the question which phase fraction is liable for the occurrence of superconductivity in SiC:B. This work is currently under way.

\section{Acknowledgments}
We acknowledge fruitful discussions with S.~Yonezawa.

This work was supported by the 21st century COE programs, ''High-Tech Research Center Project for Private Universities: matching fund subsidy'', as well as ''Diversity and Universality of Physics'' from the Ministry of Education, Culture, Sports, Science, and Technology (MEXT), and by a Grand-in-Aid for Scientific Research on Priority Area from MEXT. MK is supported by MEXT.


\begin{thebibliography}{10}
\bibitem{nVmolcomment}
{The values of the charge-carrier concentration $n$ and the molar volume
  ${V}_{\rm mol}$ used in this paper are room-temperature values. {W}e assume,
  that both quantities are only weakly temperature dependent.}
\bibitem{kGLcomment}
The Ginzburg-Landau parameter is usually defined as $\kappa_{\rm
  GL}=\lambda(0)/\xi$ with $\xi$ given by Eq.\,\ref{cohlength2}. Here we use
  the clean-limit definition $\kappa_{\rm GL}=0.96 \lambda(0)/\xi(0)$ with
  $\xi(0)$ defined by Eq.\,\ref{cohlength1}.\cite{tinkham96} The question of
  the cleanness of the sample used is discussed later in the text.
\end{thebibliography}

\end{document}